# Ultranarrow-linewidth self-injection-locked tunable blue GaN DFB laser

**Georges Perin[1,*], Laurent Lablonde[2], Marco Rossetti[3], Marco Malinverni[3], Antonino Castliglia[3], Marcus Duelk[3], Stéphane Trebaol[1]**

[1]*Univ Rennes, CNRS, Institut FOTON – UMR 6082, 6 Rue de Kerampont CS 80518, F-22305 Lannion, France*
[2]*Exail, rue Paul Sabatier, 22300 Lannion, France*
[3]*Indie Technologies Switzerland AG, EXALOS, Wagistrasse 21, CH-8952 Schlieren, Switzerland*
*[georges.perin@univ-rennes.fr](mailto:georges.perin@univ-rennes.fr)

**In this work, we demonstrate an ultra-narrow linewidth self-injection locked distributed feedback (DFB) diode laser emitting at 452 nm, achieving an intrinsic linewidth of 170 Hz with a fiber output power of 11 mW. The linewidth reduction of the DFB laser is obtained thanks to the coupling with an external cavity based on a fiber Bragg grating (FBG). Experimental results demonstrate excellent agreement with our theoretical modeling of the self-injection dynamics. Furthermore, the tuning capabilities of the system are characterized, with tunability achieved via current modulation, yielding a tuning efficiency of 300 MHz/mA over a continuous mode-hop-free range of 600 MHz. This architecture offers a robust pathway toward an integration within a compact package. Ultimately, such compact, stable, and frequency-tunable visible light sources are key for integrated optical atomic clocks and underwater lidar application.**

Narrow linewidth semiconductor lasers are fundamental building blocks for the development of next-generation, compact photonic systems. Driven by the demands of the telecommunications industry, compact narrow-linewidth laser diodes operating in the telecom C-band have achieved a high level of maturity. However, translating these advancements to the visible spectral range remains a significant challenge currently being pursued by several research groups [1–3]. The development of narrow-linewidth, tunable visible lasers is critical for a variety of applications, including laser cooling, optical atomic clocks [4], and gas sensing [5], where precise addressing of atomic transitions strictly requires narrow optical linewidths.
Additionally, visible light sources are important for underwater Light Detection and Ranging (LiDAR) [6] due to the optical transparency window of water in the blue–green spectral region [7], while system performance is governed by the laser linewidth and its frequency tunability.

Currently, commercially available lasers that meet these requirements typically rely on external cavities utilizing bulk[8] diffraction gratings, which inherently sacrifices the system's compactness. Recent research has explored more compact alternatives, such as external optical feedback from high-quality-factor microresonators [9–11], integrated photonic circuits [12–14], or fiber Bragg gratings [15]. Feedback from high quality resonators has demonstrated impressive performance, achieving linewidths as narrow as 1 kHz [9]. Despite these performances, the stability of state-of-the-art optical clocks remains fundamentally limited by the frequency noise of the interrogation lasers [4,16]. While single-mode distributed feedback laser diodes operating in the visible range have recently become commercially available, their typical free-running linewidths, in the order of hundreds of kilohertz [17], remain too broad for demanding quantum technology applications.

To bridge this gap, this work combines the single-mode operation of visible DFB laser diodes with an external cavity of a higher quality factor to drastically suppress the laser frequency noise and achieve sub-kHz linewidths. An FBG was selected for the external cavity architecture because its fiber nature enables easy integration by deporting the laser signal. Furthermore, the inherent thermal stability of FBGs compared to the laser diode allows the external cavity to operate reliably without active temperature control [18].

In this Letter, we demonstrate a 452 nm self-injection locked DFB laser (SIL) achieving an intrinsic linewidth of 170 Hz through FBG-based optical feedback. A comprehensive experimental characterization of the laser frequency noise spectrum is presented and compared with the corresponding theoretical model. Furthermore, the laser frequency tunability is investigated by modulating the current. A tuning coefficient of 300 MHz/mA and a total tuning range of 600 MHz are obtained, and the experimental results are compared with a theoretical model describing the self-injected laser tuning behavior.
The DFB GaN laser diode is realized using conventional wafer processing techniques and in the form of narrow-stripe ridge-waveguide devices working at 452.4 nm. An anti-reflection coating is deposited on the output mirror ($R_2$) reducing its reflectivity to 0.5%. This enables higher output power (up to 80 mW) and enhances sensitivity of the DFB laser to self-injection locking. The laser wavelength can be tuned via the injection current at a rate of 4 pm/mA and via temperature at a rate of 15 pm/K. Further details on the structure are given in Rossetti *et al.*[19].

Figure 1 display the experimental setup for the self-injected laser. The TO-can laser diode emission is coupled inside a fiber containing the FBG using a lens doublet (with a focal length of 4mm), the first lens is used to collimate the highly divergent beam, and the second lens to focus back the beam in the fiber core. An open-loop 3-axis PZT stage (Thorlabs, MAX312D) is used to align the FBG with the laser, allowing a coupling efficiency ($\eta$) of around 20 %. The FBG is etched into the very tip of the fiber, allowing for a short external cavity of 2 cm limited by the lens size and focal distances.

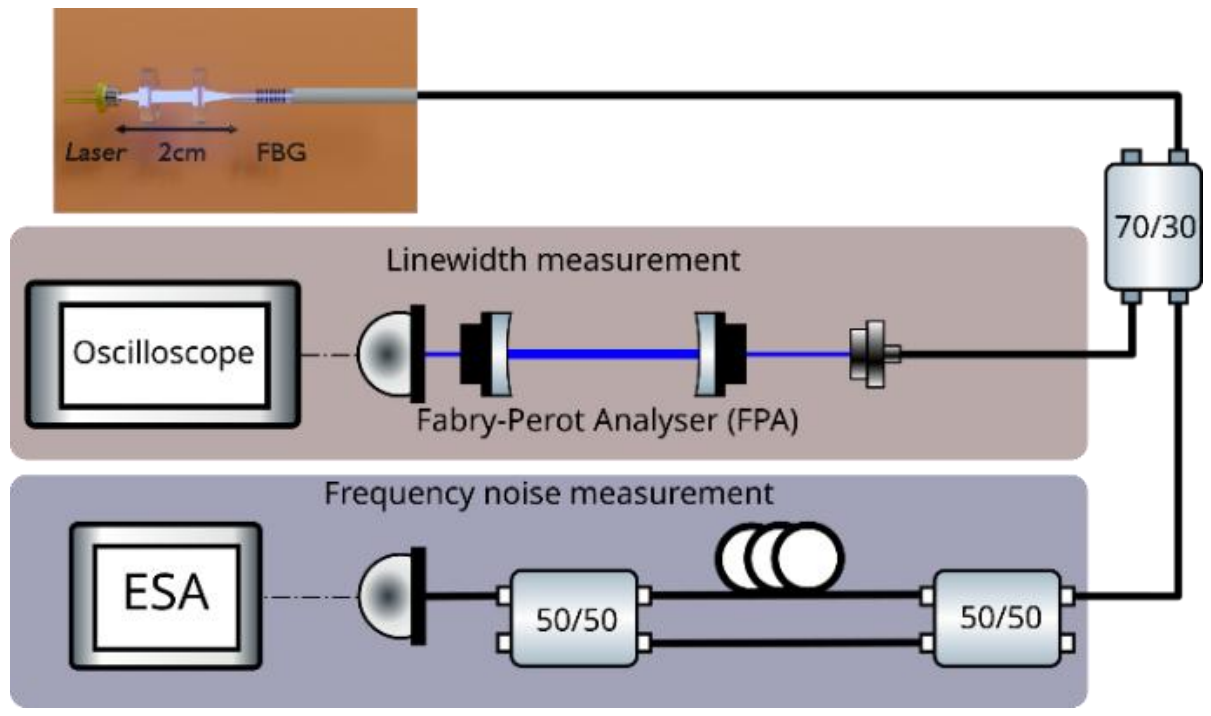

**Fig. 1.** Experimental setup of the fiber grating laser and the characterization set-up for the linewidth and the frequency noise

The output of the FBG fiber is connected to a coupler, sending 30% to a Fabry-Perot analyzer (FPA). The remaining 70% of the light is directed either to a frequency noise measurement setup or to an optical spectrum analyzer, enabling the two measurements to be performed in parallel. The frequency noise (FN) measurement setup is an auto-homodyne setup based on a Mach-Zehnder (MZ) interferometer [20].

Figure 2(a) illustrates the physical mechanism underlying the linewidth-narrowing process. Optimal self-injection locking is achieved when the Bragg reflection peak (orange) and the DFB mode (blue) are spectrally aligned to an external cavity mode (yellow) corresponding to the zero-detuning configuration.

To realize this condition, the laser current and temperature are carefully adjusted to position the emission wavelength at the maximum of the Bragg reflection spectrum.
The external cavity of the self-injection-locked DFB laser is formed by a fiber Bragg grating (FBG) with a full width at half maximum of 25 pm and a peak reflectivity of 60% at 452.45 nm. Figure 2(b) shows the Bragg transmission spectrum of a germanosilicate single-mode fiber (S405HP, orange curve) overlaid with the DFB laser emission spectrum (blue curve), demonstrating spectral alignment suitable for efficient optical feedback via the fibered grating, enabling significant linewidth reduction potentially down to the sub-kilohertz regime.

**Linewidth narrowing.** The performance of the DFB laser is first evaluated by replacing the fiber Bragg grating with a standard S405 XP single-mode fiber, and introducing a 40 dB optical isolator between the lenses to avoid optical feedback from the fiber facet. A laser linewidth of approximately 20 MHz was measured using a Fabry-Perot analyzer as illustrated in Fig. 2(c). To have a better understanding of the DBF laser spectral behavior, the FN power spectral density is measured (blue curve in Fig. 2(f)). This measurement allows to characterized the FN from 100 Hz to 1 MHz and to evaluate the intrinsic and integrated linewidth. The spectrum exhibits a usual frequency noise profile for a laser, characterized by a $\frac{1}{f}$ flicker noise slopes up to approximately 700 kHz, after which it flattens to a white noise floor of $4 \times 10^5$ Hz²/Hz. This white noise level corresponds to an intrinsic linewidth of 1.2 MHz, comparable to reported performance in the literature [17, 19].

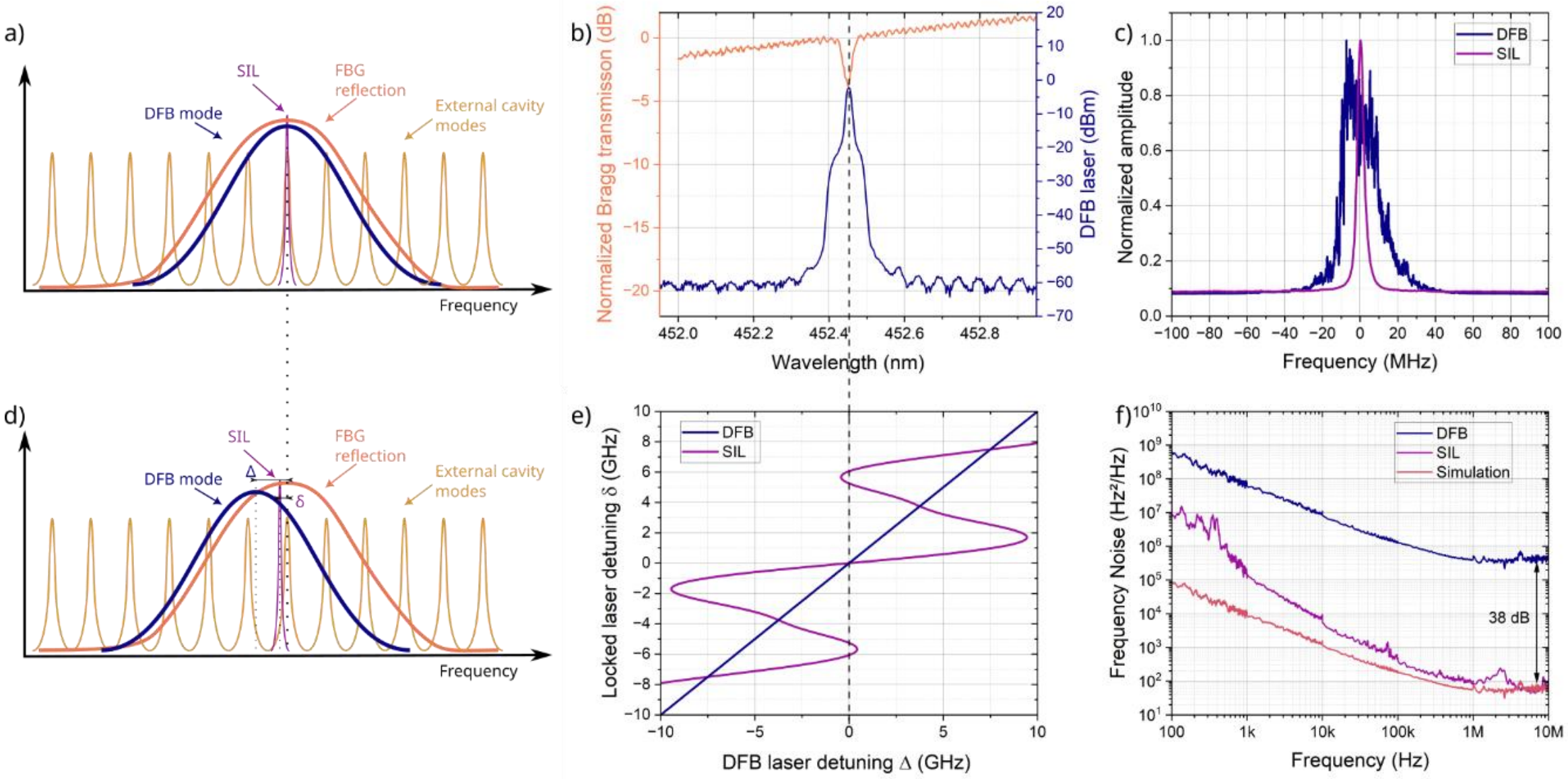

**Fig. 2.** a) Representation of the physical mechanism of the SIL for an ideal case where the DFB mode, the external cavity modes and the Bragg transmission are align. b) Comparison between the laser emission wavelength (blue curve) and the fiber Bragg grating (orange curve). c) Comparison between the linewidth of the DFB laser (blue curve) and the SIL laser (purple curve) measured at the Fabry Perot analyzer. d, e) Effect of a detuning (Δ) between the DFB mode and an external cavity mode on the detuning ($\delta$) between the SIL and an external cavity mode. f) Frequency noise comparison between the DFB laser (blue curve), the SIL laser (purple curve) and the simulated noise reduction (pink curve obtained from Eq.(1)).

The SIL laser configuration is properly established by controlling the distance between the DFB laser and the fiber, thereby tuning the external cavity round-trip phase and aligning the external cavity modes. This spectral alignment enhances optical feedback, leading to improved stability and a significant reduction of the laser linewidth from 20 MHz to 4 MHz under self-injection locking, as shown in Fig. 2(c). This value is limited by the resolution of the FPA and requires frequency-noise measurements for an accurate determination of the SIL laser linewidth.

The frequency noise of the SIL (purple curve, Fig. 2(f)) shows an important reduction in comparison to the DFB laser FN (blue curve), with the white noise dropping from $4 \times 10^5$ to 55 Hz/Hz², corresponding to a reduction of 38 dB. This impressive noise reduction corresponds to a SIL intrinsic linewidth of 170 Hz. This linewidth is nearly one order of magnitude below the state-of-the-art of narrow linewidth semiconductor laser architectures operating in the blue spectral range (Fig. 3).

The linewidth narrowing factor of the SIL can be model following the Peterman's model [21] describing the interactions between the laser emission and a coupled external cavity. In this approach, the linewidth of the SIL can be described by the following equation:

$$\Delta\nu = \frac{\Delta\nu_0}{(1+C)^2} \quad 1$$

With $\Delta\nu$ the linewidth of the SIL laser and $\Delta\nu_0$ the linewidth of the DFB laser, $C$ is the reduction coefficient expressed as:

$$C = \frac{K_{ext}\,\tau_{ext}}{\tau_{DL}}\sqrt{1+\alpha^2} \quad 2$$

depending on the round-trip times of the laser and the external cavity respectively $\tau_{DL}$, $\tau_{ext}$ and the amplitude-frequency coupling coefficient of the laser estimated to $\alpha = 4$ . The external reflection coefficient $K_{ext}$ is defined as:

$$K_{ext} = (1 - R_2\,)\sqrt{\frac{(\eta^2\,R_3)}{R_2}} \quad 3$$

With $R_2$, $R_3$ the reflectivity of the laser output facet and of the Bragg grating respectively. The coupling coefficient η determines the coupling strength between the laser and the external cavity. The theoretical reduction coefficient $(1+C)^2$ is calculated to be around 7000. The simulation of the laser frequency noise using this value is given by the pink curve of Fig. 1(f) where an excellent agreement with the experimental data is demonstrated for the white noise at high frequencies. This suggest that higher linewidth reduction is possible by changing the external cavity parameter such as increasing the Bragg reflection.

This agreement diverges at lower frequency, this can be attributed to the mechanical sensitivity of the external cavity. Indeed, the integrated linewidth can also be estimated from the frequency noise [22]. The resulting linewidths shows a reduction from 3 MHz (for the DFB laser) to 100 kHz (for the SIL laser) far from the 38 dB of reduction observed for the intrinsic linewidth. Improved mechanical stability is therefore expected to substantially reduce the low-frequency noise. In particular, integrating the external cavity into a single package, combined with bonding components and an enhanced thermal stabilization, should lead to a significant decrease in the frequency noise of the SIL in the low-frequency regime. Keeping in mind that the lower frequency noise can also be reduce by implementing a locking scheme[20].

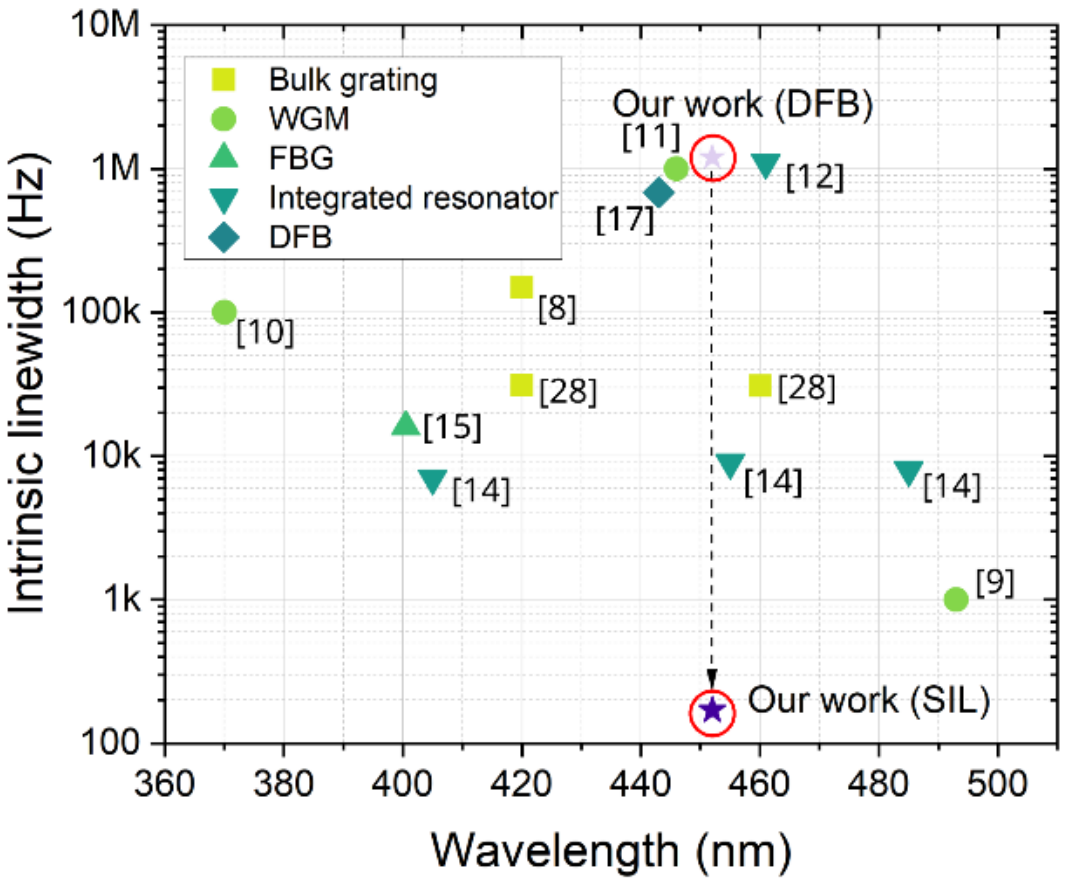

**Fig. 3.** Comparison of intrinsic linewidth of different narrow-linewidth semiconductor laser architectures in the 360-500 nm range.

**Laser tunability.** Laser tunability is essential for various applications, such as frequency locking to specific atomic transitions as well as frequency-modulated continuous-wave (FMCW) LiDAR, where performance scales with the tuning speed and agility of the laser [23].

The SIL laser tunability has been extensively investigated [24, 25], predominantly under strong optical feedback. In this configuration, when the DFB free-running frequency is detuned from the external cavity mode from Δ (Fig. 2 (d)), the self-injected signal imposes an additional phase condition that shifts the steady-state oscillation away from the free-running frequency toward the external-cavity mode. This shift, known as frequency pulling, reflects the balance between the DFB free-running frequency and the frequency selectivity of the external cavity. The frequency detuning of the SIL relative to the external cavity mode $\delta$ can be extracted from [25]:

$$\Delta = \delta + \frac{|B(\delta)|}{\tau_{DL}}\sqrt{1+\alpha^2}\sin\left(\psi_0 - \arg[B(\delta)]\right) \quad 4$$

with $\psi_0$the initial phase, $B(\delta)$ and $R_{2,eff}$ given by:

$$B(\delta) = \frac{\sqrt{R_2} - \sqrt{R_{2,eff}}e^{2\pi\delta\tau_{ex}}}{1 - \sqrt{R_2 R_{2,eff}}e^{2\pi\delta\tau_{ex}}}, \quad 5$$

and

$$R_{2,eff} = \eta^2 R_3. \quad 6$$

The SIL laser frequency tunability respect to the DFB laser detuning frequency is plotted on Fig. 2(e).

The SIL frequency is equal to the DFB frequency when detuning is minimized. But, as long as the detuning remains within the locking range, the SIL follows the external-cavity resonance and maintains self-injection locking, with the emitted frequency continuously pulled along the dispersion curve of the FBG mode. In practice, the SIL frequency lies between the free-running DFB frequency and the Bragg resonance. From Fig. 1(e), the slope of the tuning curve, K, can be extracted. This parameter corresponds to the ratio of the DFB laser tunability to that of the SIL. The experimentally measured value of K is 11.3, in good agreement with the value of 9.3 extracted from the simulated curve in Fig. 1(e).

To characterize the frequency tunability of the laser, a current ramp was applied on the DFB laser, and the resulting optical signal was sent through a Mach–Zehnder interferometer (MZI) with a 10 m delay. The laser frequency shift was extracted from the phase of the MZI output [26]. Figure 4 (red curve) displays the laser response to a 1 kHz triangular current modulation. The laser response was compared with an ideal triangular waveform, and the deviation between the two signals is shown in blue. The maximum difference remains below 30 MHz, corresponding to less than 10% of the total modulation amplitude. This characterization was performed over a wide frequency range, demonstrating a quasi-linear laser response with a tuning coefficient of approximately 300 MHz/mA from 100 Hz to 1 MHz (limited by the current driver bandwidth). The measured modulation bandwidth and the difference from an ideal triangular waveform are consistent with record values reported in the literature in the telecom band [27].

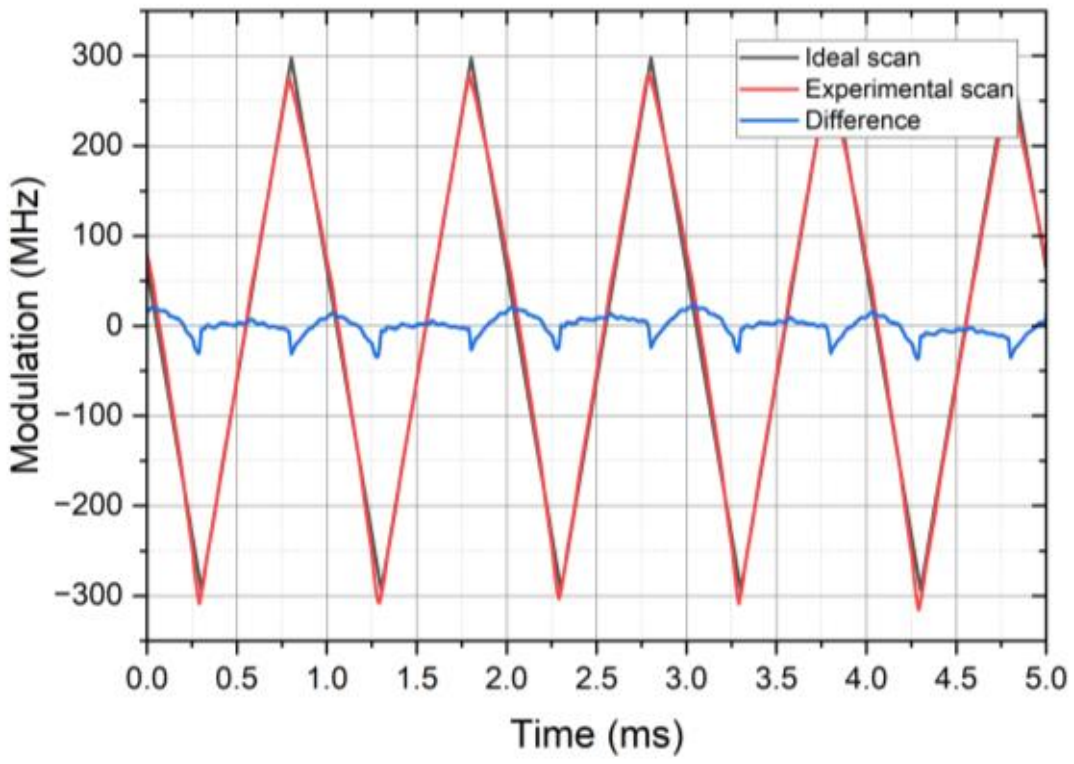

**Fig. 4.** Comparison of the laser frequency shift response to a triangular signal at 1 kHz (red curve) and a perfect triangular signal (black curve), with the residual difference between the two (blue curve).

We have demonstrated the spectral narrowing of 452 nm DFB laser by optical feedback using a fiber Bragg grating. The developed laser delivers a power of 11 mW with an intrinsic linewidth of 170 Hz, corresponding to a reduction from the DFB laser of 7000 times. To our knowledge, this performance represents the narrowest linewidth reported to date in the visible spectrum for such compact architecture. The laser also demonstrated an excellent tunability of 300 MHz/mA from 100 to 1 MHz with a depth of around 600 MHz. Furthermore, the experimental configuration is highly compatible with future integration into a robust butterfly package. This packaging is expected to significantly mitigate low-frequency noise and improve the long-term stability of the laser system, making it well suited for demanding applications in quantum technologies, atomic physics, and LiDAR.

## Funding.

I-DEMO PIA4 Bpifrance - Région Bretagne - Lannion Tregor Communauté (40898814/1, QoQeliQo). CPER PhotBreizh.

## Disclosures.

The authors declare no competing interests.

## Data Availability Statement (DAS).

The datasets generated and analysed during the current study are available from the corresponding authors upon reasonable request.